\documentstyle[12pt]{article}

\begin{document}

\hoffset = -1truecm
\voffset = -2truecm
\title{$\lambda\phi^4$ model and Higgs mass in standard 
model calculated by Gaussian effective potential approach
with a new regularization-renormalization method}
\author{Guang-jiong Ni$^1$\thanks{E-mail:
gjni@fudan.ac.cn}, Sen-yue Lou$^{2,3,1}$
\thanks{E-mail: sylou@fudan.ac.cn}, 
 Wen-fa Lu$^1$ and Ji-feng Yang$^1$ \\
\footnotesize \it $^1$Department of Physics,
Fudan University, Shanghai 200433, P. R. China\\
\footnotesize \it $^2$CCAST(World Laboratory), 
P.O. Box 8730, Beijing 100080, P. R. China\\
\footnotesize \it $^3$Institute of Modern 
Physics, Ningbo Normal College, 
Ningbo 315211, P. R. China}

\date{}
\maketitle

\begin{abstract}
Basing on new regularization-renormalization method,
the $\lambda\phi^4$ model 
used in standard model is 
studied both perturbatively and nonperturbatively (
by Gaussian effective potential). The invariant property of two mass 
scales is stressed and the existence of a (Landau) 
pole is emphasized. 
Then after coupling with the SU(2)$\times$U(1) gauge fields, the 
Higgs mass in standard model (SM) can be calculated as
$m_H\approx$138GeV. The critical temperature ($T_c$) for restoration 
of symmetry of Higgs field, the critical  energy scale ($\mu_c$, the 
maximum energy scale under which the lower excitation sector of the 
GEP is valid) and the maximum energy scale ($\mu_{max}$, at which the 
symmetry of the Higgs field is restored) in the standard model are 
$T_c\approx$476 GeV, $\mu_c\approx 0.547\times 10^{15}$Gev 
and $\mu_{\max}\approx 0.873 \times 10^{15}$ Gev respectively.
\end{abstract}

\newpage

\section{Introduction}

Year after year, the standard model (SM) in particle physics enjoys
its great success, especially after the discovery of the top quark 
in 1995$^{[1,2]}$.
Now the careful phenomenological analysis even leads to a very 
impressive conclusion that the only unobserved particle mass, the 
Higgs mass 
$m_H$, is constrained within a rather narrow interval,
say $130 \sim 150$Gev by present experimental data$^{[3,4,5]}$.

On the other hand, the calculation on $m_H$ by quantum field theory 
(QFT) 
lagged behind the experimental progress. It is well 
known that at tree level, the ratio of
$m_H^2$ to $m_W^2$ reads
\begin{equation}
\frac{m_H^2}{m_W^2}=\frac43\frac\lambda{g^2}.
\end{equation}
However, unlike the gauge coupling constant $g$, the value of 
$\lambda$
is unknown. So one has to resort to QFT beyond tree level.
Then the calculation turns out to be rather difficult and 
confused due to the divergence, counter-term and the ambiguity 
between bare and physical parameters.
Furthermore, a puzzle of so called ``triviality" existed in
$\lambda\phi^4$ model$^{[6]}$ which rendered the situation more 
complicated. 
For many years, only a lower bound and/or 
an upper bound on $m_H$ were obtained$^{[7]}$.

Nine years ago, believing in triviality and introducing a large but 
fixed 
cut off $\Lambda$, we had attacked this problem by the Gaussian
effective potential (GEP) method in QFT. We found$^{[8,9]}$
\begin{equation}
76 Gev<m_H<170 Gev
\end{equation}
which was still rather unreliable and unsatisfied since we had been 
bothered
by all these difficulties mentioned above.

Now we are in a much better position to restudy the problem. Basing 
on a new 
regularization-renormalization (R-R) method first proposed by one of 
us, Yang $^{[10,11]}$, and further applied in Refs [12,13], we can 
get rid of all the annoying divergence, counter-term and bare 
parameters
so that a clearcut value of $m_H\approx138$ GeV will emerge after the 
input of  the present accurate experimental data.

The organization of this paper is as follows: In section 2, the 
$\lambda\phi^4$ model used in SM will be studied both perturbatively 
and 
nonperturbatively (by GEP method). Then in section 3, the singularity 
(Landau pole) is stressed. the running coupling constant
and renormalization group equation are  also discussed. After a  brief
summary on  $\lambda\phi^4$ model in section 4, section 5 is devoting 
to 
the calculation of Higgs mass in SM. The final section 6 contains 
the summary and discussions.

\section{$\lambda\phi^4$ model with spotaneous symmetry breaking 
(SSB)}

The Lagrangian of $\lambda\phi^4$ with wrong sign in mass term reads
\begin{equation}
{\cal L}=\partial^{\mu}\phi
\partial_{\mu}\phi +\frac{1}{2}\sigma\phi^2
-\frac{\lambda}{4!}\phi^4.
\end{equation}
  
\subsection{One loop (L=1) calculation$^3$} 

Besides the tree level (L=0) contribution to the effective potential 
(EP)
\begin{equation}   
V_0=-\frac12\sigma\phi^2+\frac1{4!}\lambda\phi^4,
 \end{equation}
the one-loop contribution to EP is evaluated as$^6$
\begin{equation} 
V_1(\phi)=\frac12\int\frac{d^4k_E}{(2\pi)^4}
\ln(k_E^2-\sigma+\frac12\lambda \phi^2),
\qquad (<\phi>\rightarrow \phi).
\end{equation}
Denoting $M^2=-\sigma+\frac12\lambda\phi^2$,
we get 
$\frac{\partial^3V_1}{\partial 
(M^2)^3}=\frac1{32\pi^2M^2}
$ and 
\begin{equation}
V_1(\phi)=
+\frac1{32\pi^2}\left\{\left(\frac12
\lambda\phi^2-\sigma
\right)^2\left[\frac12\ln\left(\frac{
\lambda\phi^2-2\sigma}{2\mu_1^2}\right)
-\frac34\right]+C_2\left(\frac12\lambda
\phi^2-\sigma\right)+C_3\right\}
\end{equation}
with three arbitrary constants 
$\mu_1,\ C_2,$ and $C_3$. To fix them, we calculate
$\frac{dV_{eff}}{d\phi}=\frac{d}{d\phi}(V_0+V_1)=0
$. As the symmetric phase $\phi_0=0$ 
is not interesting to us, we manage to 
fix the spontaneous symmetry breaking (SSB) phase at
\begin{equation}
\phi^2_1=\frac{6\sigma}{\lambda}
\end{equation}
as that at tree level by choosing
$ \mu^2_1=C_2=2\sigma $.
At the same time, the mass square of excitation at SSB phase reads
\begin{equation}
m_\sigma^2\equiv\left.\frac{d^2V_{eff}}{d\phi^2}
\right|_{\phi=\phi_1}=2\sigma
\end{equation}
while the renormalized coupling constant is modified:
\begin{equation}
\lambda_R=
\left.\frac{d^4V_{eff}}{d\phi^4}
\right|_{\phi=\phi_1}=\lambda\left(1+\frac{
9\lambda}{32\pi^2}\right).
\end{equation}

\subsection{GEP method}

In GEP method,
we begin from a Gaussian wave functional (GWF)
$|\Psi>^{[8,9,14]}$
\begin{equation}
|\Psi>=N_f\left\{-\frac1{2\hbar}\int_{x,y}(\phi_x
-\Phi_x)f_{xy}(\phi_y-\Phi_y)\right\}
\end{equation}
with $ \int_x=\int d^3x, \ 
\Phi_x=<\phi_x>=<\Psi|\phi(x)|\Psi>$, 
etc and a correlation function 
of quantum fluctuation:
\begin{equation}
f_{xy}=f(\stackrel\rightarrow x
-\stackrel\rightarrow y)=\int\frac{d^3p}{(2\pi)^3} f_p
\exp\left[\frac i\hbar \stackrel\rightarrow p\cdot
(\stackrel\rightarrow x-\stackrel\rightarrow y)\right].
\end{equation}
Calculating the energy of system in this GWF, we get 
$E=<\Psi|H|\Psi>$ as a function of $\Phi$ and $f_p$,
A variation $\frac{\delta E}{\delta f_p}=0$
leads to 
($p=|\stackrel \rightarrow p|,\ \hbar=1)$
\begin{equation}
f_p=\sqrt{p^2+\mu^2},\qquad
\mu^2=-\sigma+\frac\lambda 2\Phi^2
+\frac\lambda 4 f_{xx}^{-1}:
\end{equation}
where $f_{xy}^{-1}$ is the inverse of $f_{xy}$,
$\int_yf_{xy}f_{yz}^{-1}=\delta^3
(\stackrel\rightarrow x-\stackrel\rightarrow z)$.
The energy $E$ is a function of 
$\phi(\equiv\Phi)$ and $\mu^2$:
\begin{equation}
E(\phi,\mu)=-\frac12\sigma\phi^2
+\frac\lambda{24}\phi^4
+\frac12I_0-\frac12(\sigma+\mu^2)I_1+\frac18\lambda
\phi^2I_1+\frac1{32}\lambda I_1^2.
\end{equation}
with
$I_0(\mu^2)\equiv f_{xx}
\equiv I_0$, $I_1(\mu^2)\equiv f_{xx}^{-1}
\equiv I_1$,
\begin{equation}
I_0=\int\frac{d^3p}{(2\pi)^3}
\sqrt{p^2+\mu^2},\qquad
I_1=\int\frac{d^3p}{(2\pi)^3}
\frac1{\sqrt{p^2+\mu^2}}.
\end{equation}

The variational condition
$\frac{\partial E}{\partial\mu^2}=0
$ leads again to $\mu^2$ equation (10) with a common factor
$I_2(\mu^2)\equiv I_2\equiv
-2\frac{\partial I_1}{\partial \mu^2}$
ignored ($I_2=0$ leads to a trivial 
sector $V_G\rightarrow V_0$ see below).
Then the GEP is defined as a function of one variable $\phi$:
\begin{equation}
V_G(\phi)\equiv E(\phi,\mu(\phi))
=-\frac12\sigma\phi^2
+\frac\lambda{24}\phi^4
+\frac12I_0-\frac1{32}\lambda I_1^2
\end{equation}
with 
\begin{equation}
\mu^2=-\sigma+\frac12\lambda\phi^2+\frac\lambda 
4I_1,\qquad \frac{d\mu^2}{d\phi}=
\frac{8\lambda\phi}{8+\lambda I_2}.
\end{equation}
Note that $I_0,\ I_1$ and $I_2$ are all divergent. 
After handling them by our new 
$R-R$ method, we obtain
\begin{eqnarray}
&&I_{2}(\mu^2) = -\frac{1}{4\pi^2}
\ln\frac{\mu^2}{\mu^2_{s}} \qquad
I_{1}(\mu^2) =
\frac{1}{8\pi^2}\mu^2(\ln\frac{\mu^2}
{\mu^2_{s}}-1)+C_{2} \\
&&I_{0}(\mu^2) =
\frac{1}{32\pi^2}\mu^4
(\ln\frac{\mu^2}{\mu^2_{s}}-
\frac{3}{2})+\frac{1}{2}C_{2}\mu^2+C_{3}
\end{eqnarray}
with $\mu^2_{s}$, $C_{2}$ and $C_{3}$ being three arbitrary 
constants. 

For discussing the SSB, we calculate
\begin{equation}
\frac{dV_{eff}}{d\phi}=\phi[
-\sigma+\frac{\lambda}{6}\phi^2+
\frac{\lambda}{4}I_{1}(\mu^2)]
\end{equation}
Besides the symmetric phase located at
$\phi_0=0
$ the SSB phase $\phi_1$ can still be located at
\begin{equation}
\phi_{1}^2=\frac{6\sigma}{\lambda}\ 
\end{equation}
by choosing
$ \mu_{s}^2=\mu_{1}^2=
\mu^2(\phi_{1})=2\sigma, \hspace*{5mm}
C_{2}=\frac{1}{8\pi^2}\mu_{1}^2\ .
$

Meanwhile
\begin{equation}
\left.\frac{d^2V_{eff}}{d\phi^2}\right|_{\phi=\phi_{1}}=2\sigma
\end{equation}
remains the same as before. However,
\begin{equation}
\lambda_{R}\equiv
\left.\frac{d^4V_{eff}}{d\phi^4}\right|_{\phi_{1}}
=\lambda\left[1+\frac{9}{32}
\frac{\lambda}{\pi^2}+\frac{3^3}
{2^{10}}\frac{\lambda^2}{\pi^4}\right]
\end{equation}
is further modified but closed at the order of $\lambda^3$ though 
GEP method amounts to add up the loop contributions of cactus
diagram of $\lambda\phi^4$ model up to $L\rightarrow \infty$.

\section{Singularity in GEP, running coupling constant and 
renormalinzation
group equation}

Being a nonperturbative approach in QFT, GEP
method is essentially different from any perturbative 
calculation up to $L$ being fixed
large number. To see this, let us 
concentrate on gap equation (the first one of (16))
 $(\frac{\mu^2}{2\sigma}=x, 
\frac{\phi^2}{2\sigma}=y)$:
\begin{equation}
y=\left(\frac2\lambda+\frac1{16\pi^2}\right)x
-\frac1{16\pi^2}x\ln x+\left(\frac1\lambda
-\frac1{16\pi^2}\right).
\end{equation}
It is interesting to see that $y$ is a 
single-valued but not monotonic function of
$x$. The SSB phase is located at ($x=1,\ y=\frac3\lambda)$
whereas the decreasing of $x$ to the left side 
can not reach the symmetric phase $y=0$ 
at $x\rightarrow0$ as long as $\lambda <16\pi^2$. On the right
side, increasing of $x$ will lead to a maximum of $y,\
y_{\max}$,  at $x_c$
\begin{equation}
\frac{\mu_c^2}{\mu_1^2}\equiv x_c=\exp\left(
\frac{32\pi^2}{\lambda}\right),\qquad
\frac{\phi_{\max}^2}{2\sigma}\equiv y_{\max}
=\frac1{2\lambda}(x_c+1).
\end{equation}
Further running of $x$ will arrive at the remote destination 
$(x=x_{\max},\ y=0)$ where the symmetric phase 
$\phi_0=0$ is restored at high energy excitation
($\mu>\mu_c$) sector. 

Return back to $V_G$ as a function of $\phi$,
we see that $\phi_{\max}$ corresponding to $x_c$ (or 
$\mu_c$) is a singular point of $V_G$ because 
$\frac{d\mu^2}{d\phi}$ is divergent at $x_c$.
It divides $V_G$ into two branches (sectors).
The low energy excitation sector ($\mu<\mu_c$)
contains the SSB phase ($\phi=\phi_1$), whereas 
the high energy excitation sector
($\mu>\mu_c$)  contains the symmetric phase
$(\phi=0)$ with very low energy in the whole 
system.
No other stationary state exists. So for low excitation
particles, the system is staying at SSB phase
and will not collapse to the symmetric
phase ($\phi=0$) of the other sector because of 
the barrier at $\phi_{\max}$.

In GEP scheme, we define the running coupling constant
(RCC) as (${\cal J}\equiv\ln(\mu^2/(2\sigma))$
$$
\bar\lambda(\mu(\phi))\equiv\frac{d^4V_G}{d\phi^4}=
\mu^{-4}\left( 2^{21}\cdot3\lambda^3\phi^2\pi^8\mu^2
+2^{25}\lambda\pi^{10}\mu^4+2^{15}\cdot3\lambda^{5}
\phi^4\pi^6-2^{20}\lambda^4\phi^4\pi^8
\right.
$$
$$
-(2^{17}\cdot3\lambda^4\phi^2\pi^6\mu^2 +2^{21}\pi^8
\mu^4\lambda^2-2^{15}\lambda^{5}\phi^4\pi^6
){\cal J}
+(2^{11}\cdot3\lambda^5\phi^2\pi^4\mu^2
-2^{16}\lambda^3\pi^6\mu^4)
{\cal J}^2
$$
\begin{equation}
\left.
+2^{13}\lambda^4\pi^4\mu^4
{\cal J}^3
-2^5\cdot7\lambda^5\pi^2\mu^4
{\cal J}^4+2\lambda^6\mu^4{\cal J}^5
\right )\left (
2^5\pi^2-\lambda{\cal J}\right )^{-5}
\end{equation}
and a beta function
$$
\beta(\bar\lambda)\equiv\mu\frac d{d\mu}\bar\lambda(\mu)=
\frac1{\left(2^5\pi^2
-\lambda{\cal J}\right)^{6}
\mu^{4}}\left((-2^{13}\cdot15\lambda^5\pi^
{4}\mu^4+2^{12}\cdot5\lambda^6\phi^2\pi^4\mu^2)
{\cal J}^3
\right.
$$
$$
+[2^{12}\cdot15\lambda^6\phi^2\pi^4\mu^2
+2^{17}\pi^6(45\lambda^4\mu^4+\lambda^6\phi^4
-15\lambda^5\phi^2\mu^2)]
{\cal J}^2
+2^6\cdot15\lambda^6\pi^2\mu^4{\cal J}^4
$$
$$
+2^{26}\cdot15\lambda^2\pi^{10}\mu^4
-2^{27}\pi^{10}(5\lambda^3\phi^2\mu^2-\lambda^4\phi^4)
-2^{22}\pi^8(5\lambda^5\phi^4-15\lambda^4\phi^2\mu^2)
+2^{16}\cdot15\lambda^6\phi^4\pi^6
$$
\begin{equation}
\left.
+\left(2^{22}\pi^8\left(15\lambda^4\phi^2\mu^2
-2\lambda^5\phi^4-30\lambda^3\mu^4\right)
+2^{17}\pi^6\left(5\lambda^6\phi^4
-30\lambda^5\phi^2\mu^2\right)\right)
{\cal J}
\right)
\end{equation}
with
\begin{equation}
\beta(\bar\lambda)|_{\mu_1}=
\frac{3\lambda^2}{(4\pi)^2}
+\frac{135\lambda^4}{4(4\pi)^6}
\end{equation}
which can be compare with 
$\beta(\lambda) =
\frac{3\lambda^2}{(4\pi)^2}
-\frac{17\lambda^3}{3(4\pi)^4}+\cdots
$
usually quoted as in Ref. [15].

Obviously, from Eq. (25), we see that
there is a pole of five order in
$\bar\lambda $ at $\mu=\mu_c$.
On the other hand, we can define a RCC and beta function in 
one-loop calculation of EP 
\begin{equation}
\bar\lambda^{(1)}(M)=\frac{d^4V_{eff}}{d\phi^4}
=\lambda +\frac{\lambda^2}{32\pi^2}\left(3\ln
\frac{M^2}{2\sigma}+8+4\frac\sigma{M^2}-4\frac{\sigma^2}{M^4}
\right),
\end{equation}
\begin{equation}
\beta^{(1)}(\bar\lambda)=M\frac{d\bar\lambda}{dM}
=\left(3-4\frac\sigma{M^2}+8\frac{\sigma^2}{M^4}
\right)\frac{\lambda^2}{16\pi^4}
\end{equation}
with 
\begin{equation}
\beta(\bar\lambda)^{(1)}|_{\mu_1}=
\frac{3\lambda^2}{(4\pi)^2}.
\end{equation}
It is clear that there is no pole in $\bar\lambda^{(1)}(M)$.
That is because the contribution in $V_{eff}$ is a sum
of L=1 diagram while $V_G$ is a sum of L$\rightarrow\infty$
diagram. To re-find the pole from one loop EP, one
may perform an improvement
on $\bar \lambda(M)$ by 
renormalization group equation (RGE). Modifying the right
hand side of Eq. (30) 
\begin{equation}
\bar\mu\frac{d}{d\bar\mu}\bar\lambda(\bar\mu)=
\frac{3\bar\lambda(\bar\mu)^2}{(4\pi)^2}.
\end{equation}
by $\lambda\rightarrow\bar\lambda(\bar\mu)$. Integrating 
the RGE (31) yields
\begin{equation}
\bar\lambda(\bar\mu)=\frac{\lambda_R}{1-\frac{3}{16\pi^2}
\lambda_R\ln\frac {\bar\mu}{\mu_1}},
\end{equation}
where $\mu_1=\bar\mu|_{\phi_1},\ \lambda_R=\bar\lambda(\mu_1)$.

Evidently, there is a simple pole, 
$\bar\mu=\bar\mu_c=\mu_1\exp\left(
\frac{16\pi^2}{3\lambda_R}
\right)$,
the so-called Landau pole in Eq. (32). The location of 
sigularity
should be
an invariant feature of $\lambda\phi^4$ model at QFT level.
So the difference between 
$\mu_c$ and $\bar\mu_c$ implies a relation 
between two running mass scales,
$\mu$ in GEP method and $\bar\mu$ in L=1 RGE calculation.

\section{Brief summary on $\lambda\phi^4$ model}

(a) The $\lambda\phi^4$ model is well defined at classical level 
by Lagrangian shown at Eq. (3). However, it is not well
defined at QFT level by ${\cal L}$ solely before it is supplemented
by three constants:
$C_1=-\ln\mu_1,\ C_2$ and $C_3$.

(b) While $C_3$ is trivial (it only affects the whole shift of EP), 
fixing $\mu_1$ and $C_2$ is equivalent to reconfirming two mass 
scales, 
$m_\sigma^2=2\sigma$ and $\phi^2_1=\frac{6\sigma}\lambda, $ in
$\lambda\phi^4$ model at QFT level.

(c) Now we understand that the \it invariant \rm
meaning of parameter $\lambda$ in ${\cal L}$ is not  
a coupling constant but the ratio of these two mass squares, 
$m_\sigma^2/\phi_1^2=\lambda/3$, at any order of loop (L) expansion 
theory even at GEP (L$\rightarrow\infty$) level.

(d) The prominent difference between perturbative theory
(L=finite) and nonperturbative theory (L$\rightarrow
\infty$) like GEP method (or RGE) lies in the fact 
that in the latter case there is a singularity at 
GEP, the critical mass scale
$\mu_c=\mu_1\exp\frac{16\pi^2}\lambda$ (or Landau pole,
$\bar\mu_c=\mu_1\exp\frac{16\pi^2}{3\lambda_R}$, in RGE)
whereas in the former there is no
singularity.
An elementary example is the  geometric series
$S_n=1+r+\cdots +r^n$ is analytic whereas 
$S_n|_{n\rightarrow \infty}=\frac1{1-r}$
$(|r|<1)$ has a pole.

(e) Formally, when the mass of a physical particle 
exceeds a critical value, $\mu>\mu_c$, a phase transition 
is triggered. The system would collapse to symmetric phase, 
$\phi_0=0$. Safely speaking, $\mu=\mu_c$ is the upper bound in 
energy scale of $\lambda\phi^4$ model at QFT level with
SSB.

(f) In summary, at QFT level, $\lambda
\phi^4$ model with SSB is characterized by two mass 
scale: $(\phi_1^2=6\sigma/\lambda$ and $\mu_1^2
=2\sigma)$ and one singularity in $\mu^2$,
$\mu_c^2=\mu_1^2\exp (32\pi^2/\lambda)$. It is
a renormalizable, nontrivial and effective
(up to critical energy $\mu_c$) theory.

\section{Calculation of Higgs mass}

We are now well prepared to calculate the Higgs mass $m_H$
in SM by GEP method. 

As ageneralization of Eq. (10), we start from a GWF:
$$
|\Psi>=\exp\left\{-\frac12
\int_{xy}\{[\xi(x)-\bar\xi(x)]F_{xy}(\bar\xi)
[\xi(y)-\bar\xi(y)]-[W^\mu(x)F_{xy}(\bar W)
W^*_\mu(y)\right.
$$
\begin{equation}
\left.\ +W^*_\mu(x)F_{xy}(\bar W)W^\mu(y)]
-Z^\mu(x)F_{xy}(\bar Z)Z_\mu(y)+A_\mu(x)
F^{\mu\nu}_{xy}(\bar A)A^\mu(y)\}\right\},
\end{equation}
where $\xi$ is the real Higgs 
field while $W^\mu,\ Z^\mu$ and $A^\mu$
are fields of $W,\ Z$ bosons and photon respectively, 
$\bar \xi=<\Psi|\xi|\Psi>$ etc..
The quantum fluctuation correlation function
\begin{equation}
F_{xy}(\bar B)=C_B^3\int\frac{d^3p}{(2\pi)^3}
\sqrt{p^2+\mu_B^2}\exp[iC_B
\stackrel\rightarrow p\dot (\stackrel\rightarrow 
x-\stackrel\rightarrow y)],
\end{equation}
($\bar B=\bar W,\ \bar Z,\ \bar A,\ \bar\xi,\ 
C_A=C_W=C_Z=\sqrt{3/2}\equiv C^{1/3},\ C_\xi=1$),
is controlled by the mass parameter $\mu_B$ which 
is determined via variational procedure
and is different for different fields 
($\bar\xi\rightarrow \xi$ again):
\begin{eqnarray}
\mu_{\xi}^{2} &=& -
\sigma+\frac{\lambda}{2}\left[\xi^{2}
+\frac{1}{2}I_{1}(\mu_{\xi}^{2})\right]
+\frac{3}{4}g^{2}CI_{1}(\mu_{W}^{2})
+\frac{3}{8}C(g^{2}+{g'}^{2})
I_{1}(\mu_{Z}^{2}), \\
\mu_{W}^{2} &=& 
\frac{1}{4}g^{2}\left[\xi^{2}+\frac{1}{2}
I_{1}(\mu_{\xi}^{2})\right]
+g^{2}CI_{1}(\mu_{W}^{2})+
\frac{g^{4}C}{g^{2}+{g'}^{2}}I_{1}(\mu_{Z}^{2})
+\frac{g^{2}{g'}^{2}C}
{g^{2}+{g'}^{2}}I_{1}(\mu_{A}^{2}=0), \\
\mu_{Z}^{2} &=& 
\frac{1}{4}(g^{2}+{g'}^{2})\left[
\xi^{2}+\frac{1}{2}I_{1}(\mu_{\xi}^{2})\right]
+2\frac{g^{4}C}{g^{2}+{g'}^{2}}I_{1}(\mu_{W}^{2}),
\end{eqnarray}
where $g$ and $g'$ are coupling constants in SU(2) $\times$ U(1)
gauge model. Some explanations are
important:\\
(a) We need not introduce any counter terms related to gauge fields 
for
ensuring the massless property of gauge bosons at symmetric phase
($\xi$=0) because the low energy symmetric phase is not at the same
sector with the SSB phase ($\xi=\xi_{1}$) under consideration as 
shown in
pure $\lambda\phi^4$ theory ($\lambda<16\pi^{2}$). \\
(b) While $\mu^{2}_{W}|_{\xi_{1}}=m^{2}_{W}$ and
$\mu^{2}_{Z}|_{\xi_{1}}=m^{2}_{Z}$ are the
observed mass square of W and Z bosons at SSB phase, the parameter
\begin{equation}
\mu_{\xi}^{2}|_{\xi_{1}}
\equiv\mu_{1}^{2}=\frac{\lambda}{3}\xi_{1}^{2}\neq 2\sigma
\end{equation}
is not the Higgs mass square as that at tree level. We will soon find
the expression for Higgs mass after the quantum corrections are
added.\\
(c) We set the mass parameter for photon field, $\mu_A$, always
zero, $\mu_A$=0, as shown in the last term of Eq. (36). \\
(d) After performing the same renormalization procedure as in 
$\lambda\phi^4$ model, 
$I_{1}(\mu^{2})$ in Eqs. (35$\sim$37) has the form
\begin{equation}
I_{1}(\mu^{2}) =
\frac{1}{8\pi^2}\mu^{2}\left(\ln\frac{\mu^{2}}
{\mu^{2}_{1}}-1\right)+C_{2}.
\end{equation}
Taking $C_{2}=\frac{\mu_{1}^{2}}{8\pi^{2}}=I_{1}(0)$ further, we have
$I_{1}(\mu^{2}_{1})=0$.
Basing on Eqs. (35)-(37) and using the experimental data$^{[3,4,5]}$:
\begin{eqnarray}
&& \alpha^{-1}=\frac{4\pi}{g^2\sin^2\theta}=128.89,\qquad
\sin^{2}\theta\equiv\frac{{g'}^{2}}{g^{2}
+{g'}^{2}}=0.2317,\nonumber \\
&& m_{W}=80.359 {\rm Gev}, \hspace*{5mm}
m_{Z}=91.1884 {\rm Gev},
\end{eqnarray}
we manage to find the values of $\mu_1$, $\lambda$ and $\sigma$.
Denoting
\begin{equation}
w_{1}=\frac{m_{W}^{2}}{\mu_{1}^{2}}, \hspace*{5mm}
a=\frac{m_{Z}^{2}}{m_{W}^{2}}=1.2877
\end{equation}
and calculating $(36)|_{\xi_{1}}
\times (g^{2}+{g'}^{2})-
(37)|_{\xi_{1}}\times
g^{2}$,
one obtains
\begin{eqnarray}
& &\{\sec^{2}\theta-a
-\frac{g^{2}C}{8\pi^{2}}[a(\ln a-1)-
\sec^{2}\theta
+2\cos^{2}\theta]\}w_{1}-\frac{g^{2}C}
{4\pi^{2}}(\sec^{2}\theta
-\cos^{2}\theta) \nonumber \\
&=& \frac{g^{2}C}{8\pi^{2}}
(\sec^{2}\theta-2\cos^{2}\theta+a)
w_{1}\ln w_{1}
\end{eqnarray}
or
\begin{eqnarray*}
0.0210070w_{1}-0.0104422 &=& 
0.0103063w_{1}\ln w_{1}
\end{eqnarray*}
One finds (apart from a meaningless solution $w_1 \gg 1$, see final
discussion)
\begin{equation}
w_{1}=\frac{m_{W}^{2}}{\mu_{1}^{2}}=0.3183153
\end{equation}

Substituting the value (43) into Eq. (37)$|_{\xi_{1}}$ with (38), one 
finds:
\begin{equation}
\lambda=1.0139453
\end{equation}
Then it is easy to find the value of $\sigma$ from 
Eq.(35)$|_{\xi_{1}}$:
\begin{equation}
\frac{\sigma}{\mu_{1}^{2}}=0.5034030
\end{equation}
which means that $\mu^2_1$ is not far from its value at tree level,
2$\sigma$.

Apart from the fermion contribution to be added below, the Higgs mass 
square reads (see Ref. [9])
\begin{equation}
\left. \frac{d^{2}V_{eff}}{d\xi^{2}}
\right|_{\xi_{1}}= \frac{2\xi_{1}^{2}}{3}
\frac{[4(\lambda r+3s)-\lambda(\lambda
r+s)I_{2}(\mu_{1}^{2})]}{[8r+(\lambda r
+s)I_{2}(\mu_{1}^{2})]}  
=\mu_{1}^{2}\left(1+\frac{3}{\lambda}\frac{s}{r}\right)
\end{equation}
where $I_2(\mu^2)=-\frac1{4\pi^2}\ln\frac{\mu^2}{2\sigma}$,
\begin{eqnarray}
r &=&
2+g^{2}CI_{2}(\mu_{W}^{2})-g^{8}C^{2}
I_{2}(\mu_{W}^{2})I_{2}
(\mu_{Z}^{2})/(g^{2}+{g'}^{2})^{2} \nonumber \\
s &=& \frac{3}{8}(g^{2}+{g'}^{2})^{2}
\left[\left(\frac{g^{4}}
{(g^{2}+{g'}^{2})^{2}}-
\frac{1}{4}\right)\right. 
g^{2}C^{2}I_{2}(\mu_{W}^{2})
I_{2}(\mu_{Z}^{2}) \nonumber 
\\
& & \left.  -\frac{g^{4}C}{
(g^{2}+{g'}^{2})^{2}}
I_{2}(\mu_{W}^{2})-
\frac{C}{2}I_{2}(\mu_{Z}^{2})\right]
\end{eqnarray}

We see that the quantum fluctuation effect of gauge fields on the 
Higgs mass is very small: $\frac{3}{\lambda}\frac{s}{r}=-0.00853833 $.

Moreover, the fermions will contribute to Higgs mass at one loop level
as discussed in Ref. 12. A fermion with mass $m_i$ contributes:
$-\frac{{G_{i}}^{2}}{2\pi^{2}}m_{i}^{2}
\ln\frac{m_{i}^{2}}{\mu_{1}^{2}}$.
In SSB theory of SM, $G_{i}=(\frac{m_{i}}{\xi_{1}})$,
so the Higgs mass $m_{H}$
should be evaluated as
\begin{equation}
(\frac{m_{H}}{\mu_{1}})^{2}
=1+\frac{3}{\lambda}\frac{s}{r}
-\frac{1}{2\pi^{2}}\frac{\lambda}
{3}\sum_{i=e,\mu,\tau}
(\frac{m_{i}}{\mu_{1}})^{4}
\ln(\frac{m_{i}}{\mu_{1}})^{2}
-\frac{3}{2\pi^{2}}\frac{\lambda}
{3}\sum_{q=u,d,s,c,b,t}
(\frac{m_{q}}{\mu_{1}})^{4}
\ln(\frac{m_{q}}{\mu_{1}})^{2}
\end{equation}
The extra factor 3 for quarks comes from the color freedom. Actually,
only the top quark with $m_t$=175GeV makes the main contribution.
Eventually, we find
\begin{equation}
(\frac{m_{H}}{\mu_{1}})^{2}=0.943251\qquad {\rm i.e.,}\qquad
m_{H}=138.331{\rm Gev}\approx 138 {\rm Gev}\ .
\end{equation}

\section{Summary and discussions}

(1) The motivation of adopting GEP method is the following. The value 
of
weak mixing (Weinberg) angle $\theta$ derived from the experiments of 
neutral current process
\begin{equation}
\sin^{2}\theta=\frac{{g'}^{2}}
{g^{2}+{g'}^{2}}\approx 0.2317
\end{equation}
is different from the value derived from the mass ratio of W, Z 
bosons,
\begin{equation}
1-\frac{m_{W}^{2}}{m_{Z}^{2}}\approx 0.2234
\end{equation}
due to the quantum corrections to all orders in
perturbation theory. However, to evaluate the discrepancy 3.7\%
in
QFT is not easy. Then GEP method has the advantage of providing an
analytically calculable scheme. As shown in Eqs. (35$\sim$37), we
assume that the gauge fields undergo the same quantum fluctuation as 
that
of Higgs field in a GWF but with different mass parameters, which
are
linking together. Hence the difference between (50) and (51) provides 
a
possibility to find the value of $\lambda$ and thus the Higgs mass.

(2) It is interesting to see that 
$\lambda\approx 1.$ Then we can find 
a critical temperature T$_c$ for restoration of SSB
phase ($\xi_1$) to symmetric phase ($\xi=0$), as 
discussed in Ref. [16] or Ref. [12]:
\begin{equation}
T_{c}=\sqrt{\frac{12}{\lambda}}m_{H}.
\end{equation}
Substituting the value of $\lambda$ and $m_H$ here, we find
\begin{equation}
T_{c}=475.886{\rm GeV}\approx 476{\rm Gev}
\end{equation}
which is not far from 510 GeV as estimated in Ref. 12 by other method.

(3) In pure $\lambda\phi^4$ model, there is a critical mass scale 
$\mu_c(=\sqrt{2\sigma}\exp(16\pi^2/\lambda)$, beyond which the system 
will collapse to symmetric phase. After coupling with gauge fields,
this critical value $\mu_\xi=\mu_c$ should be solved from Eqs. 
(35$\sim$37)
together with the vanishing condition of denominator of 
$\frac{\partial\mu_\xi^2}{\partial \xi}$, i.e.,
\begin{equation}
8r+(\lambda r+s)I_2(\mu_\xi^2)=0.
\end{equation}
Numerical calculation yields
\begin{equation}
\mu_c\approx 0.547\times 10^{15} {\rm Gev}.
\end{equation}
Furthermore, the maximum energy scale, 
$\mu_{\max}=
\mu_{\xi}|_{\xi=0}$, as can be solved from Eqs. 
(5$\sim$7) with $\xi=0$ is approximately:
\begin {equation}
\mu_{\max}\approx 0.873\times 10^{15} {\rm Gev}
\end{equation}
at which the symmetry of Higgs field is restored in 
high energy sector ($\mu_\xi>\mu_c$) whereas at $T=T_c$ 
symmetry restoration occurs at low excitation sector
($\mu_\xi<\mu_c$).

(4) The advantage of our new R-R method can be
seen as follows. In Ref. [9], 
$I_2(\mu_1)\sim \frac1{2\pi^2}\ln\frac\Lambda{\mu_1}$
was logarithmically divergent whereas now 
$I_2(\mu_1^2)=0.$ So whole calculation becomes quite clear and well
under control.

(5) As stressed in $\lambda\phi^4$ theory, the model is characterized
by two mass scales $(\xi_1=\sqrt{6\sigma/\lambda}$ and 
$\mu_1=\sqrt{2\sigma})$ and one singularity 
$\mu_c=\sqrt{2\sigma}\exp(16\pi^2/\lambda)$). After coupling with 
gauge fields, both $\xi_1$ and $\mu_1$ are modified to some extent
while keeping their ratio form $\xi_1/\mu_1=\sqrt{3/\lambda}$
invariant. The critical value $\mu_c$ is strongly suppressed to
$\mu_c\approx 0.547\times 10^{15}$ Gev, which could be viewed as 
the upper bound of energy scale in SM with SSB.

Nonetheless, the whole model is well defined (reconfirmed) 
at nonperturbative QFT level.

(6) Because the experimental data are not quite fixed yet$^{[5]}$,
for checking the sensitivity of 
our results to input, we have calculated the value of $m_H$
over a wide rage: $80.26{\rm Gev}<m_w<80.36{\rm Gev}\ 
(m_Z=91.1884\ {\rm Gev})$ and $0.2316<\sin^2\theta<0.2325$. 
The result shows that the average value is 
$$ <m_H>\approx 140.96 {\rm Gev}$$
while $m_H^{\max}\approx 143.11 {\rm Gev}$ and 
$m_H^{\min}\approx 124.92 {\rm Gev}$.

(7) Finally, we would like to compare some recent literatures on 
Higgs mass. The estimation by Altarelli and G. Isidori$^{[17]}$ 
or by Eillis et al$^{[18]}$ is not far from that of ours. On the 
other hand, the  prediction of $m_H\sim 2{\rm Tev}$ as in Ref. [19]
seems too high to be considered.

We thank Profs. Su-qing Chen, Y-s Duan, T. Huang, 
H. C. Lee, K. Wu, H. L. Yu, X-m Zhang, Z-x Zhang, Z-y Zhu and Dr.
H-q Zheng for discussion and encouragement. This work
was supported in part by the National Science Foundation in China.

\newpage
\small
\leftline{\large\bf References} 
 \begin{enumerate}
\item 
 CDF collaboration, F. Abe et al., Phys. Rev. Lett. {\bf 
74},
2626 (1995). 
\item 
DO Collaboration, S. Abachi et al., Phys. Rev. lett. {\bf
74}, 2632 (1995).
\item 
P. B. Renton, Invited talk given at the 17th 
International
symposium on Lepton-Photon Interactions, 10-15/8 1995, Beijing, China,
CERN-PPE/96-63, 13.05.96.
\item 
G. Altarelli, status of precision tests of the standard model, 
CERN-TH/96-265, \\ hep-ph/9611239.
\item J. Gillies, Warsaw Conference, CERN COURIER 36, (7), 1 (1996).
\item D. J. E. Callaway, Phys. Rep. 167, 241 (1988).
\item M. Sher, Phys. Rep. 79, 273 (1989).
\item 
G-j Ni, S-y Lou and S-q Chen, Phys. Lett. B {\bf 200}, 161
(1988).
\item 
 S-y Lou and G-j Ni, Phys. Rev. D {\bf 40}, 3040 (1989).
\item
J-f Yang, Thesis for PhD, Fudan Uni., 1994, Unpublished; 
Report No. hep-th/9708104.
\item J-f Yang and G-j Ni, Acta Physica Sinica, 4, 88 (1995);
 G-j Ni and J-f Yang, Phys. Lett. B393, 79 (1997).
\item 
G-j Ni and S-q Chen, preprint, (1996); Report No. hep-th/9708155,
 accepted by Acta Physica Sinica (overseas version).
\item 
G-j Ni and H-b Wang, preprint, (1997); Report No. hep-th/9708457.
\item     
 G-j Ni, S-y Lou and S-q Chen, Int. J. Mod. Phys. A {\bf 
3},
1735 (1988).
\item  C. Itzykson and J. B. Zuber, {\it Quantum Field Theory},
McGraw-Hill Inc. p. 653 (1980).
\item L. Dolan and R. Jackiw, Phys. Rev. D9, 3320 (1974).
\item G. Altarelli and G. Isidori, Phys. B337, 141 (1994).
\item J. Eillis, G. L. Fogli and E. Lisi, Indications
from Precision Electroweak Physics confront Theoretical 
Bounds on the Mass of the Higgs Boson, CERN-Th/96-216,
LBNL-39237, hep-ph/9608329.
\item 
R. Ibanz-Meier and P. M. Stevenson, Phys. Lett. B297,144 (1992);\\
M. Consoli, Phys. Lett. B 305, 78 (1993);\\
V. Branchina, M. Consoli and N. M. Stivala, Z. Phys. C., 57, 251 
(1993);\\
K. Halpern and K. Huang, Phys. Rev. Lett., 74, 3526 (1995).
 \end{enumerate}

\end{document}